\documentclass[slac_one]{revtex4}
\usepackage{graphicx}
\usepackage{fancyhdr,color}
\definecolor{Blue}{named}{Blue}
\definecolor{Red}{named}{Red}
\definecolor{Green}{named}{ForestGreen}
\definecolor{Black}{named}{Black}
\definecolor{Olive}{named}{OliveGreen}

\begin{document}

\title{{\small{2005 International Linear Collider Workshop - Stanford,
U.S.A.}} \hfill {\small\rm CERN-PH-TH/2005-142}\\ 
\vspace{12pt}
News from polarized $\boldmath{e^-}$ and $\boldmath{e^+}$ 
at the ILC} 

%

\author{Gudrid Moortgat-Pick}
\affiliation{TH Division, Physics Department, CERN, CH-1211 Geneva 23, 
Switzerland}
%


\begin{abstract}
The proposed International Linear Collider (ILC)
is well-suited for discovering physics
beyond the Standard Model and for precisely
unravelling the structure of the
underlying physics. The physics return of the ILC can be maximized by
the use of polarized beams, in particular the simultaneous polarization of the $e^-$ 
and the $e^+$ beam.
Ongoing physics studies are accompanied by active R\&D on the machine part
for generating polarized beams and for 
measuring the polarization with high precision at the ILC.
Some new results on the physics case and on the 
technical aspects of the polarization of both beams 
are briefly summarized.
\end{abstract}

\maketitle

\thispagestyle{fancy}


\section{INTRODUCTION \\[-.5em]}   
It is well accepted that
beam polarization will play an important role in the programme of the 
International Linear Collider (ILC). 
The polarization of the electron beam is foreseen for the baseline
design~\cite{scope}. A high degree of at least 80\% polarization is
envisaged, but new results indicate that even 90\% should be
achievable. A polarized electron beam would already provide a valuable
tool for measuring precisely Standard Model (SM) processes and for diagnosing 
candidates of new physics.

Polarizing simultaneously the electron and the positron beam is
currently discussed as an upgrade possibility for the ILC.  In the
report of the polarization working group POWER (POlarization at Work
in Energetic Reactions) \cite{Moortgat-Pick:2005cw}, it is shown that
the full potential of the linear collider could be realized only with
the polarization of both the $e^-$ and $e^+$ beams. In addition to
high-precision studies of the SM and detailed analyses of the
properties of new particles and of new kinds of interactions, the
polarization of both beams would also enable indirect searches with
high sensitivity for new physics in a widely model-independent
approach.  Consequently, very active R$\&$D is currently ongoing for
all beam polarization issues: polarized $e^{\pm}$ sources,
polarization transport, polarization measurement, as well as
reliability aspects.  In the following a short summary is given about
the news from this interesting field; all physics examples can be
found in \cite{Moortgat-Pick:2005cw}.  For more details see also
other contributions in these proceedings~\cite{lcws05}.

\section{NEWS FROM THE PHYSICS CASE FOR POLARIZED $\boldmath{e^-}$ 
AND $\boldmath{e^+}$\\[-.5em]}
A main task of future experiments in high-energy physics will not only
be to discover physics beyond the SM but also to
reveal the structure of the underlying physics and to determine
the model precisely.  It is expected that the clean signatures and in
particular the precise measurements made possible by a high-luminosity
linear collider at a known and tunable beam energy are perfectly
suited to complement and extend all kinds of new physics discoveries
that will be 
made at the Large Hadron Collider (LHC), which is scheduled to start
in 2007.

\subsection{Direct searches for physics beyond the SM\\[-.5em]}

One of the best motivated extensions of the SM is Supersymmetry
(SUSY).  This theory predicts new SUSY particles that carry the same
quantum numbers as their SM partner particles, with the exception of
the spin, which differs by half a unit. Prominent examples are the
scalar particles, the selectrons/spositrons $\tilde{e}^{\pm}_{\rm L,R}$,
which have to be associated to their SM partners, the left- and
right-chiral electrons/positrons.

To test such assumptions, 
the pairs {\color{Red}$\tilde{e}^+_{\rm L}\tilde{e}^-_{\rm R}$}
produced only in the $t$-channel process
must be experimentally separated from the pair
{\color{Blue}$\tilde{e}^+_{\rm R}\tilde{e}^-_{\rm R}$} produced also in 
the $s$-channel. It has been shown that even a
highly polarized electron beam will not be sufficient to separate the pairs, 
since both are produced with almost
identical cross sections and have the same decay.
Applying simultaneously polarized positrons, the pairs get 
different cross sections,
can be isolated, and the $\tilde{e}_{L}^+$ and $\tilde{e}^-_R$ can be
identified by charge separation;
see Fig.~1 (left panel)~\cite{Moortgat-Pick:2005cw}.

As another consequence of SUSY, the SU(2) and U(1) SUSY Yukawa
couplings have to be identical to the corresponding SM gauge
couplings.  Assuming that the masses and mixing parameters of the
neutralinos have been predetermined in the gaugino/higgsino sector,
the production cross sections of $\tilde{e}_{\rm R}^+\tilde{e}_{\rm R}^-$ 
and $\tilde{e}_{\rm L}^+ \tilde{e}^-_{\rm R}$ can be exploited to
derive the Yukawa couplings.  
However, in the case where the two pairs have almost identical cross sections 
and decay modes, $\tilde{e}_{\rm R,L}^\pm \to e^\pm
\tilde{\chi}^0_1$, the different combinations of
$\tilde{e}_{\rm R}$ and $\tilde{e}_{\rm L}$ can only be distinguished
by the initial beam polarization of both beams, see Fig.~1 
(right panel)~\cite{Moortgat-Pick:2005cw}.
\begin{figure}[htb]
\begin{minipage}{10cm}
\hspace{-3cm}
\setlength{\unitlength}{1cm}
\begin{picture}(6,6.5)
\setlength{\unitlength}{1cm}
\put(-1.1,3){\mbox{\includegraphics[height=.14\textheight]{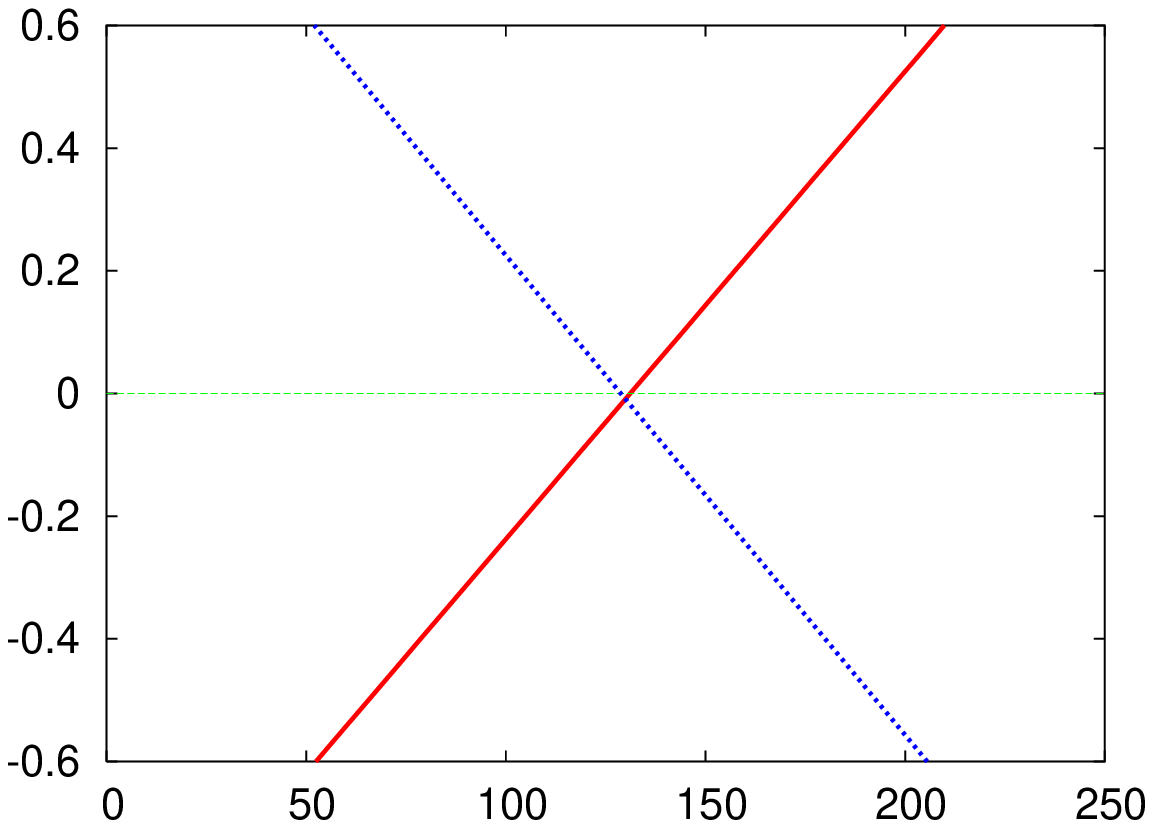}}}
\put(-.6,3.){\mbox{\includegraphics[height=.015\textheight,width=.22\textheight]{0028-box}}}
\put(-1.1,0.1){\mbox{\includegraphics[height=.14\textheight]{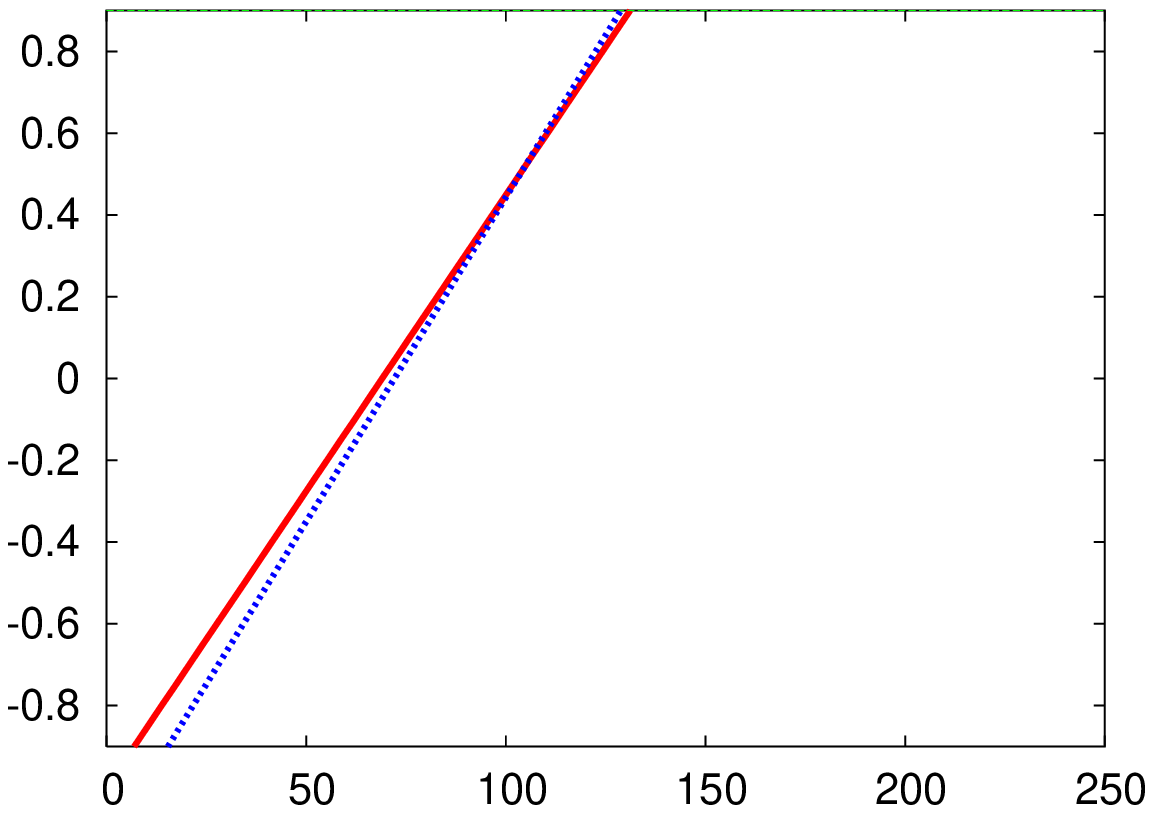}}}
\put(-1.25,4.7){\footnotesize $P_{e^+}$}
\put(-1.25,1.7){\footnotesize $P_{e^-}$}
\put(-1.4,3.1){\tiny $0.9\to$}
\put(.8,5.8){\scriptsize $P_{e^-}=+0.9$}
\put(2.4,5.4){\scriptsize\color{Red} $\tilde{e}^+_L\tilde{e}^-_R$}
\put(.0,5.4){\scriptsize\color{Blue} $\tilde{e}^+_R\tilde{e}^-_R$}
\put(1.7,2.5){\scriptsize\color{Red}$\tilde{e}^+_L\tilde{e}^-_R$}
\put(.1,2.5){\scriptsize\color{Blue} $\tilde{e}^+_R\tilde{e}^-_R$}
\put(1.43,3.7){\Large$\uparrow$}
\put(1.4,.5){\scriptsize $\sqrt{s}=500$~GeV}
\put(-.4,6.4){\scriptsize 
$e^+e^-\to\tilde{e}^+_{L,R}\tilde{e}^-_{R}\to e^+e^- \tilde{\chi}^0_1\tilde{\chi}^0_1$}
\put(.5,-.2){\scriptsize cross section [fb]}
\put(4,3.1){\includegraphics[width=.2\textheight,height=0.13\textheight]{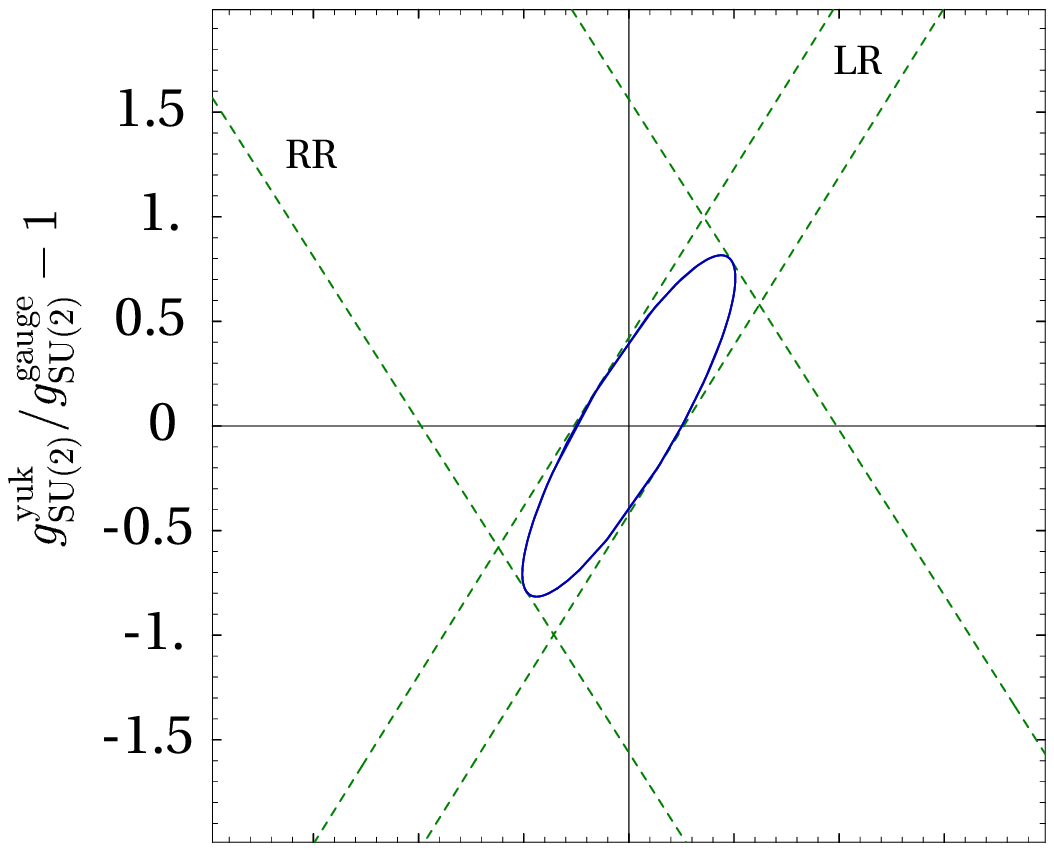}}
\put(4,0.15){\includegraphics[width=.2\textheight,height=0.13\textheight]{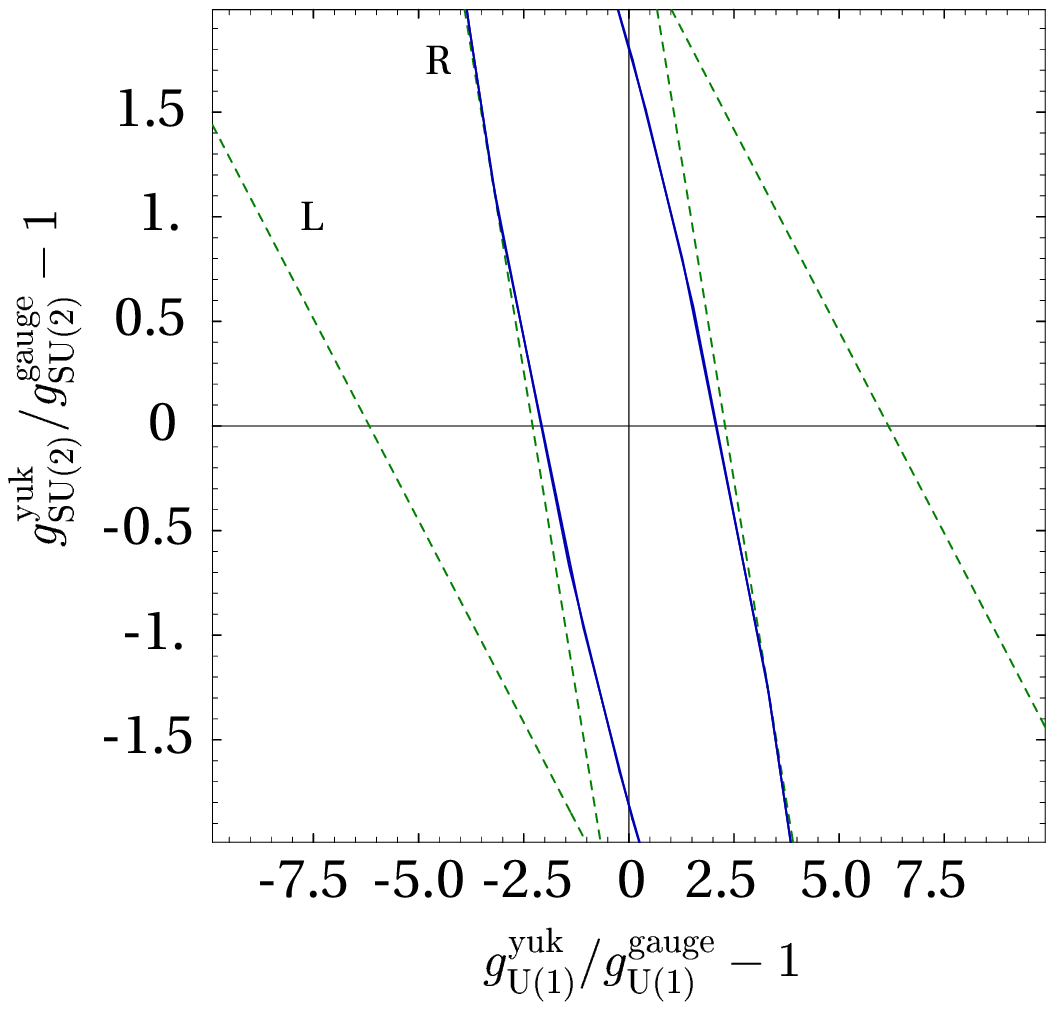}}
\put(5,6.4){\scriptsize 
$e^+e^-\to\tilde{e}^+_{L,R}\tilde{e}^-_{R}\to e^+e^- \tilde{\chi}^0_1\tilde{\chi}^0_1$}
\put(.5,-.2){\scriptsize cross section [fb]}
\end{picture}
\end{minipage}
\begin{minipage}{7.5cm}
\begin{tabular}{l}
Fig.~1: {\it Left panel: Test of chiral quantum number --}\\
{\it separation of the selectron pair
$\tilde{e}_{\rm L}^+\tilde{e}^-_{\rm R}$ is not possible}\\ 
{\it with $e^-$ polarization alone, e.g. $P_{e^-}=+90\%$}\\ 
{\it (lower plot). If, however, both beams are polarized,}\\ 
{\it the RR configuration separates the pairs $\tilde{e}_{\rm L}^+\tilde{e}^-_{\rm R}$,
 $\tilde{e}_{\rm L}^+\tilde{e}^-_{\rm L}$}\\
{\it (see arrow, upper plot).}\\
{\it Right panel: Test of Yukawa couplings --} \\
{\it 1$\sigma$ bounds on the determination of the supersymme-}\\ 
{\it tric U(1) and SU(2) Yukawa couplings between $e^+$,}\\ 
{\it $\tilde{e}_{\rm R,L}^+$ and $\tilde{\chi}^0_i$ from 
selectron cross-section measure-}\\
{\it ments; R (L) means $P_{e^-} = +90\%$ ($-90\%$) (lower}\\ 
{\it plot). In the upper plots both beams are polarized}\\ 
{\it with the values $(P_{e^+},P_{e^-}) = (-60\%,+90\%)$ (LR)}\\ 
{\it and $(+60\%,+90\%)$ (RR).}
\end{tabular}
\end{minipage}
\end{figure}

As can be seen from these examples, the availability of 
polarized positrons may be absolutely essential for the determination of
the underlying physics. 

The polarization of both beams allows a direct probe not only of the
chiral quantum numbers, as shown in Fig.~1, but also a test of the spin
of particles produced in resonances.

A prominent example is the production of a spin-0 particle, e.g.\ the
scalar neutrino in $\mu^+\mu^-$
production~\cite{Moortgat-Pick:2005cw}.  Since the sneutrino couples
only to left-handed $e^\pm$, the peak is strongest for the
configuration LL: $(P_{e^-},P_{e^+})=(-80\%,-60\%)$, which points
directly to resonance production of spin-0 particles: the SM
background is strongly suppressed and we obtain a $S/B\sim 11$, whereas
with $(P_{e^-},P_{e^+})=(-80\%,0\%)$ one obtains only $S/B\sim 4$; see
Fig.~2 (left panel).  Since 100\%-polarized beams are not possible,
the signal is still not fully suppressed in the LR configuration.
Conversely, in the case of a spin-1 resonance, e.g. from a $Z^\prime$
particle, the corresponding resonance curve would have the strongest
peak for the LR configuration with a similar polarization dependence
as the SM background; see Fig.~2 (right panel).

This simple example shows how one can directly probe the nature of the
interaction if the polarization of both beams is available.

\vspace{.1cm}
\begin{figure}[htb]
\begin{minipage}{10cm}
\hspace{-3cm}
\setlength{\unitlength}{1cm}
\begin{picture}(6,5)
\setlength{\unitlength}{1cm}
\put(-1.5,-10.3){\includegraphics[width=1.2\textwidth,height=1.85\textwidth]{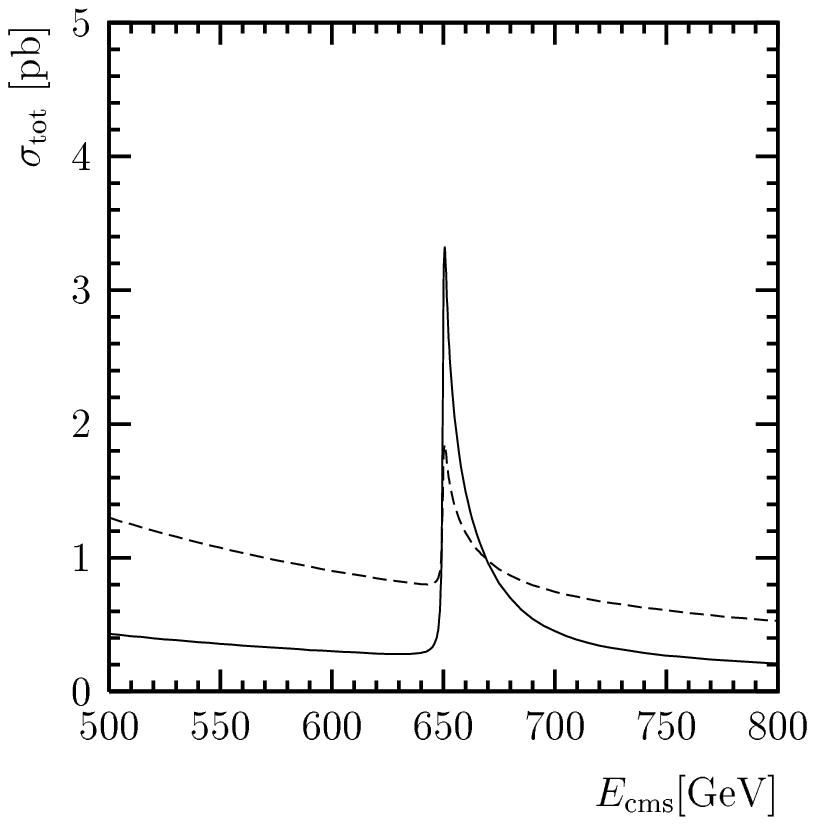}}
\put(4.5,0){\includegraphics[width=.5\textwidth,height=.5\textwidth]{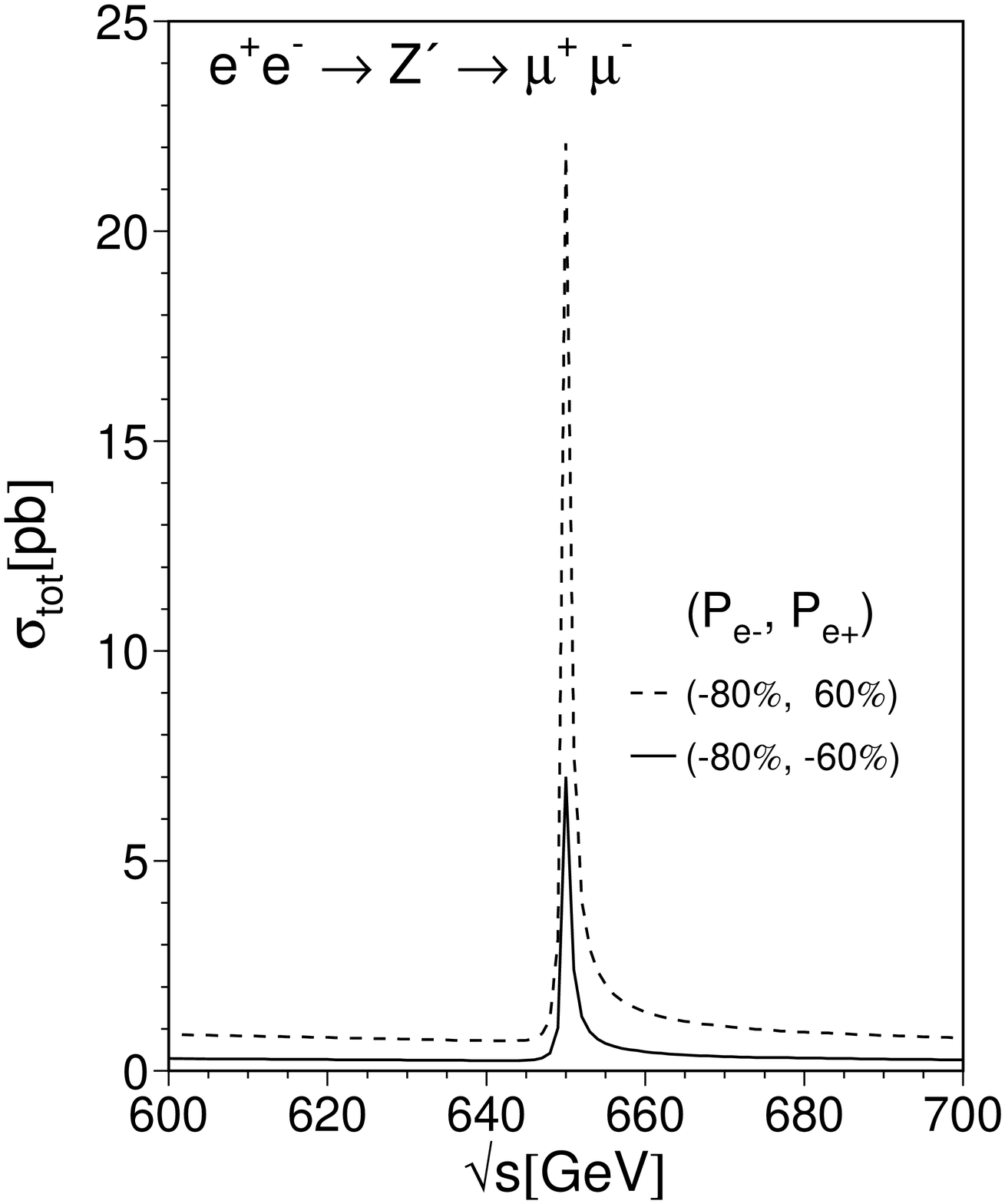}}
\put(0.3,4.6){\small $e^+e^-\to \tilde{\nu}_{\tau}\to \mu^+\mu^- $}
\put(2.5,2.5){\tiny $(P_{e^-},P_{e^+})$}
\put(2.5,2.1){\tiny $\cdot \cdot$ (-80\%,+60\%)}
\put(2.5,1.8){\tiny $-$  (-80\%,-60\%)}
\end{picture}
\end{minipage}
\begin{minipage}{7.5cm}
\begin{tabular}{l}
Fig.~2: {\it Direct probe of spin-0 in resonance production --}\\
{\it sneutrino production in the R-parity-violating model}\\ 
{\it versus $Z^\prime$ production in the SSM $Z^\prime$ model. Resonance}\\ 
{\it  production for $e^+e^-\to \tilde{\nu}_{\tau} \to \mu^+ \mu^-$  (left panel) and}\\ 
{\it for $e^+e^-\to Z^\prime \to \mu^+ \mu^-$  (right panel) for different}\\ 
{\it configurations of beam polarization:}\\ 
{\it $(P_{e^-},P_{e^+})=(-80\%,+60\%)$ (dashed),}\\ 
{\it \phantom{$(P_{e^-},P_{e^+})=$ }$(-80\%,-60\%)$ (solid).}
\end{tabular}
\end{minipage}
\end{figure}

Beam polarization is also important for mass
measurements of SUSY  particles 
in the continuum.  In many cases the worst background is
$WW$ pair production, which can be significantly reduced using
right-handed polarized $e^-$ and left-handed polarized $e^+$.
A factor of about 2.6 can be gained in the ratio $S/\sqrt{B}$
with $(P_{e^-},P_{e^+})=(+80\%,-80\%)$ compared to $(+80\%,0)$.
With
polarized $e^-$ and $e^+$ beams, the muon-energy edges, at around 65
and 220~GeV, can clearly be reconstructed.\\[1em]

\vspace*{-.4cm}
\begin{figure}[htb]
\begin{minipage}{10cm}
\begin{picture}(8,8)
\setlength{\unitlength}{1cm}
\put(-5.7,-1.4){\mbox{\includegraphics[height=.15\textheight,width=.25\textheight]
{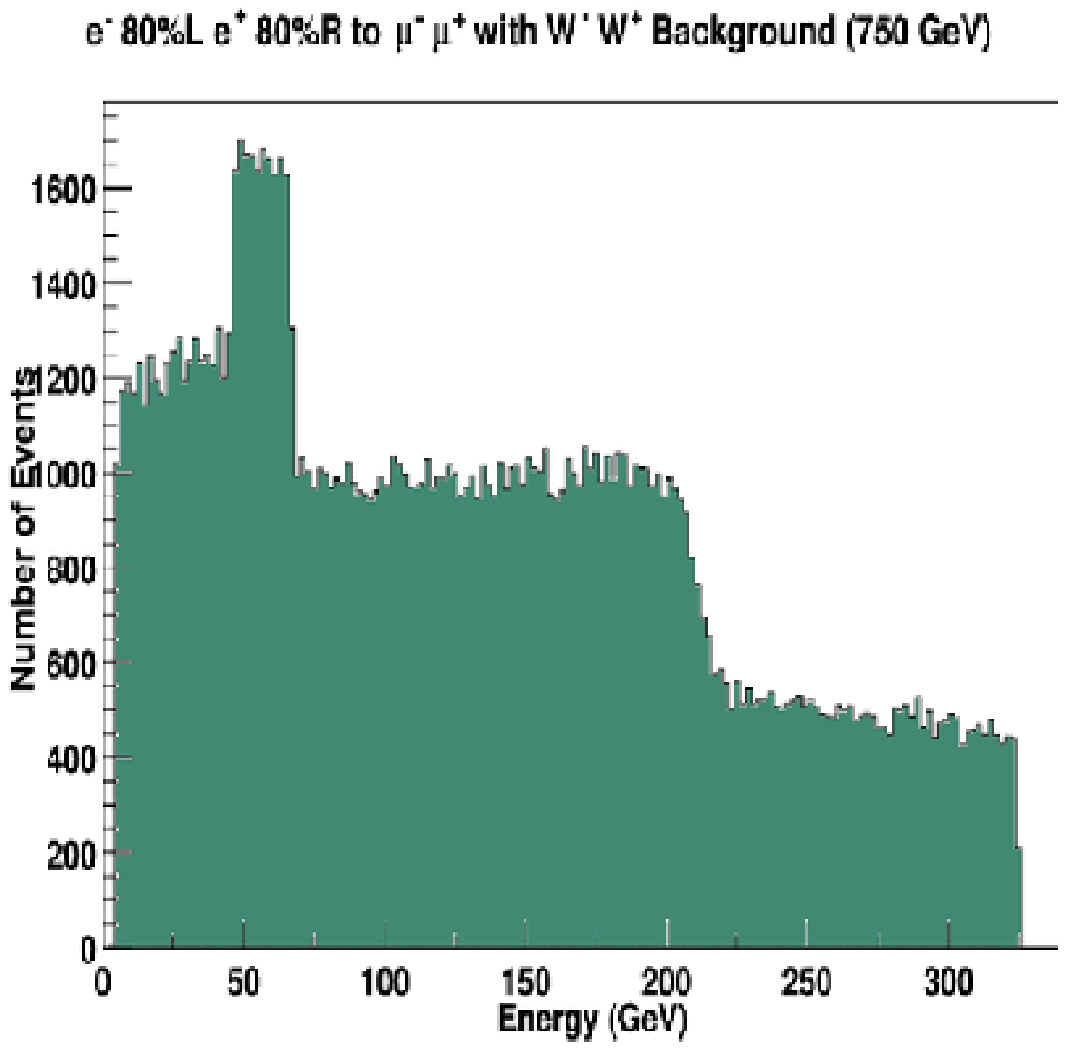}}}
\put(-5.2,1.70){\mbox{\includegraphics[height=.06\textheight,width=1.5\textheight]{0028-box}}}
\put(-5.95,-.7){\mbox{\includegraphics[height=3cm,width=.5cm]{0028-box}}}
\put(-5,-1.63){\mbox{\includegraphics[height=.5cm,width=5cm]{0028-box}}}
\put(.1,-1.4){\mbox{\includegraphics[height=.15\textheight,width=.22\textheight]
{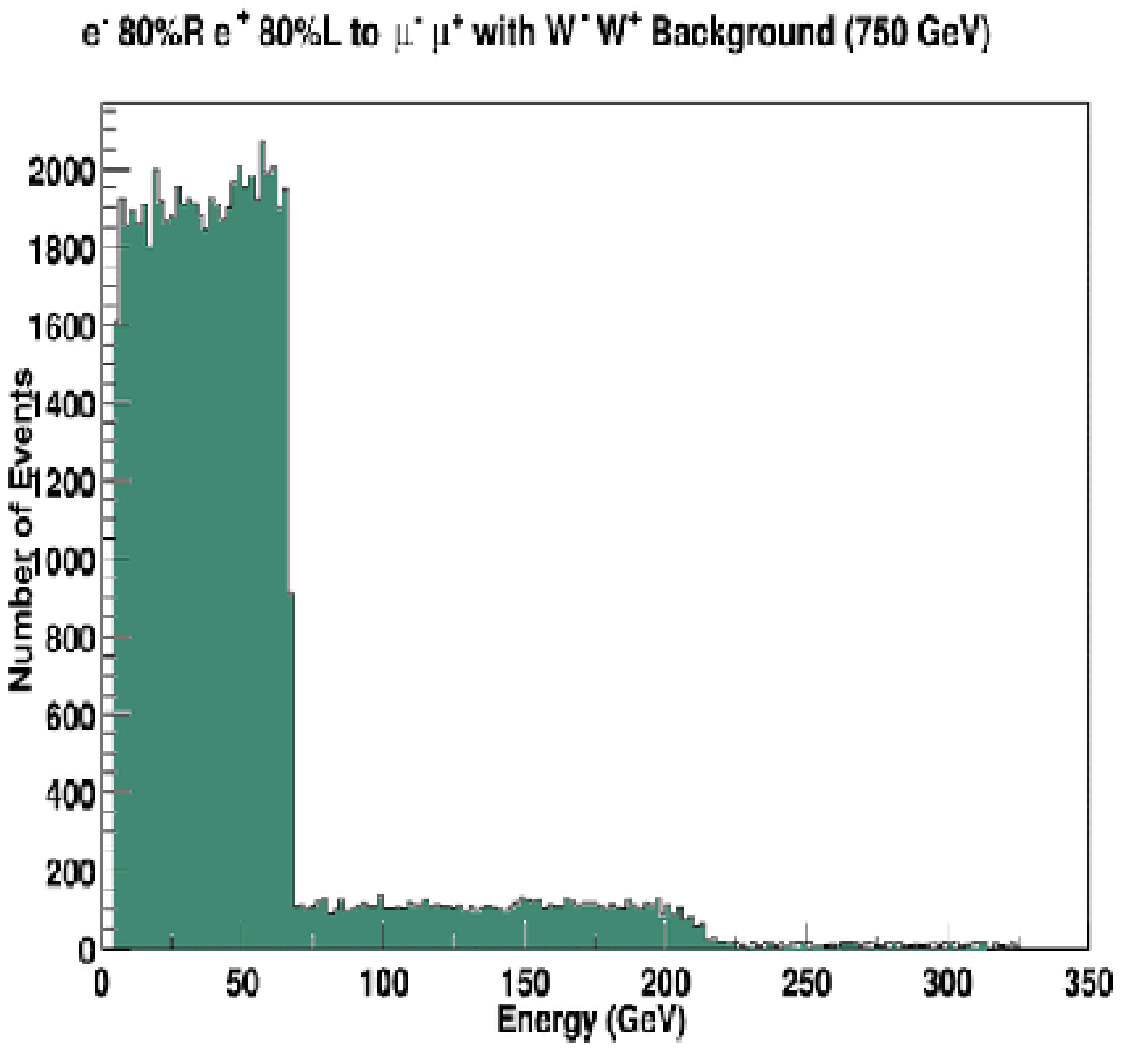}}}
\put(0,1.93){\mbox{\includegraphics[height=.04\textheight,width=.8\textheight]{0028-box}}}
\put(-.2,-.7){\mbox{\includegraphics[height=3cm,width=.5cm]{0028-box}}}
\put(-.1,-1.6){\mbox{\includegraphics[height=.5cm,width=5cm]{0028-box}}}
\put(-5.6,1.9){\tiny Number of events}
\put(.2,1.9){\tiny Number of events}
\put(-3,-1.5){\tiny Energy [GeV]}
\put(2.4,-1.5){\tiny Energy [GeV]}
\put(-1.8,1.9){\tiny $\sqrt{s}=750$~GeV}
\put(3.4,1.9){\tiny $\sqrt{s}=750$~GeV} 
\put(3.1,-.2){\footnotesize $WW$ backgr.}
\put(4,-.7){\large $\downarrow$}
\put(-1.8,0.5){\footnotesize $WW$ backgr.}
\put(-1,-.0){\large $\downarrow$}
\put(-3.8,1.4){$\leftarrow${\footnotesize $\tilde{\mu}_R$}}
\put(-2.5,1.2){\footnotesize $\tilde{\mu}_L$}
\put(-2.2,.9){$\downarrow$}

\end{picture}
\end{minipage}
\begin{minipage}{6.5cm}
\vspace{-.5cm}
\begin{tabular}{l}
Fig.~3: {\it Muon energy spectrum: $\mu^+\mu^-$ events}\\
{\it (incl. $W^+W^-$). Left panel: $W^+W^-$ background }\\
{\it is dominant for $(P_{e^-},P_{e^+})=(-80\%, +80\%)$,}\\ 
{\it so that the edges cannot clearly be reconstructed.}\\
{\it Right panel: Only with polarized $e^-$ and $e^+$ beams}\\ 
{\it both muon-energy edges, at around 65~GeV and}\\
{\it 220~GeV, can be reconstructed with}\\ 
{\it $(P_{e^-},P_{e^+})=(+80\%, -80\%)$.}\\
{\it This leads to a smuon masses determination in}\\ 
{\it the continuum up to a few GeV uncertainty.  }\\ 
\end{tabular}
\end{minipage}
\end{figure}

\vspace*{-.2cm}
Another example of new physics, where also the background suppression is important, is
the search for direct signatures of massive spin-2 gravitons.
A signature for direct graviton production, envisaged in
formulations of gravity with extra spatial dimensions, is a relatively
soft photon and missing energy.  The major background process is $\gamma \nu
\bar{\nu}$ production.
Since the neutrino coupling is only left-handed, the
background has nearly maximal polarization asymmetry and, consequently,
polarized electron and positron beams are extremely efficient in
suppressing the $\gamma\nu\bar\nu$ effects.
Compared with the case of only polarized electrons, the background process
can be suppressed by a factor of about 2, 
whereas the signal will be enhanced by a factor of about 1.5.
                                                                                
In general, such enhancements of cross sections and of the ratio $S/B$ may be
particularly important at the edge of the kinematical reach
of the machine. Enabling a look just around such a corner with 
the help of polarized positrons
may be crucial and may motivate possible upgrades.
\subsection{Indirect searches for large scales of new physics\\[-.5em]}
Some new physics scales, such as those
characterizing gravity in models with extra dimensions or the
compositeness scale of quarks and leptons, could be too large to be
directly accessible at energies of present and future 
accelerators. Therefore it will be important 
to develop strategies for indirect searches beyond the kinematical limit
for new physics. 
Thanks to the clear signatures and its high luminosity,
the ILC also has a large discovery potential in indirect searches 
in a largely model-independent approach. 

Effective contact interactions (CI) 
represent a general tool for
parametrizing at `low-energy' the effects of non-standard dynamics
characterized by exchanges of very high-mass states between the SM particles.

\begin{figure}[htb]
\setlength{\unitlength}{1cm}
\begin{minipage}{5cm}
\begin{picture}(4.0,3.7)
\put(-.7,0){\includegraphics[width=5cm,height=4cm]{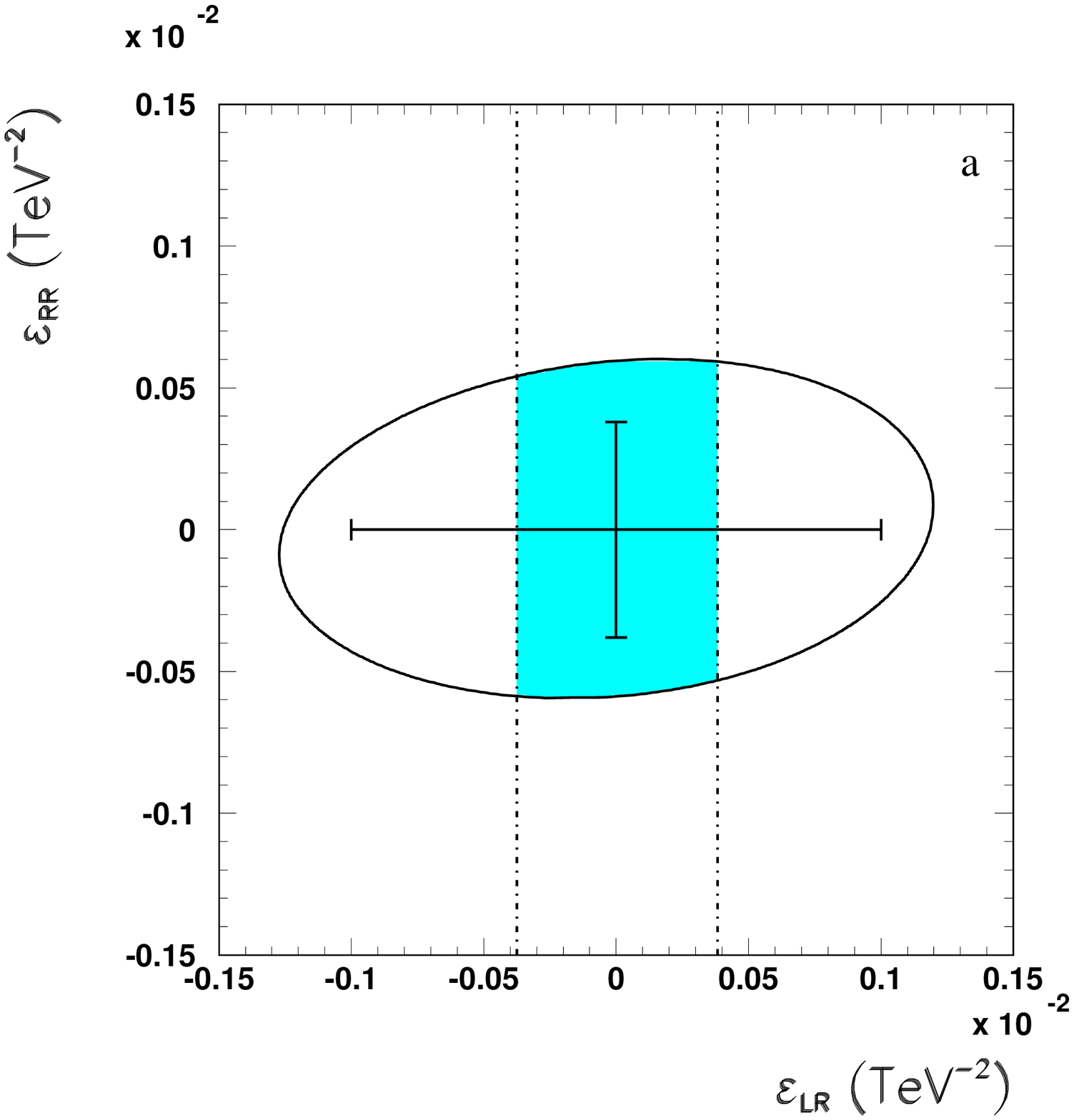}}
\end{picture}
\end{minipage}\hspace{-.3cm}
\begin{minipage}{12cm}
\begin{tabular}{l}
Fig.~4: 
{\it In Bhabha scattering the four-fermion CIs are parametrized by three}\\
{\it parameters ($\epsilon_{\rm RR}$, $\epsilon_{\rm LR}$, $\epsilon_{\rm
LL}$). The $t$-channel contributions depend only on $\epsilon_{\rm LR}$, whereas}\\ 
{\it the $s$-channel contribution depends only on pairs
($\epsilon_{\rm RR}$, $\epsilon_{\rm LR}$), ($\epsilon_{\rm LR}$,
$\epsilon_{\rm LL}$).  In order}\\ 
{\it to derive model-independent bounds it is necessary to have 
both beams polarized. Tight}\\ 
{\it bounds up to
$5\times 10^{-4}$ TeV$^{-2}$ can be derived via a $\chi^2$ test assuming that no deviations}\\
{\it from the SM
are measured in the observables $\sigma_0$, $A_{\rm FB}$, $A_{\rm LR}$ and $A_{\rm LR,FB}$ (within the}\\
{\it experimental 1~$\sigma$ uncertainty).}
\end{tabular}
\end{minipage}
\end{figure}

Extra neutral gauge bosons $Z'$ can be probed by their virtual effects on
cross sections and asymmetries.  For energies below a $Z'$ resonance, measurements of
fermion-pair production are sensitive to the ratio of $Z'$ couplings
and $Z'$ mass. Positron-beam polarization
with $(P_{e^-},P_{e^+})=(80\%,40\%)$  improves considerably, by about a factor 1.4 compared
with $P_{e^-}=80\%$ only,
the measurement of the $b\bar{b}$ couplings of $Z'$.
The crucial point is the fact that the systematic
errors can be significantly reduced when both beams are polarized.
\subsection{One more opportunity to find even tiny hints for new physics\\[-.5em]}
Extremely sensitive tests of the SM can be performed with the help of
electroweak precision observables. These can be measured with very high
accuracy at the GigaZ option of the ILC, i.e. running with high luminosity
at the $Z$-boson resonance.
Measuring accurately the
left--right asymmetry allows a determination of the effective weak mixing angle
 $\sin^2\theta_{\rm eff}$ with the highest precision.  However, in order to
exploit the gain in statistics at GigaZ, the relative uncertainties on the beam
polarization have to be kept below 0.1\%. This ultimate precision cannot be
reached with Compton polarimetry, but by using a modified Blondel scheme,
which requires the polarization of both beams.

With the polarization of both beams, using the Blondel scheme, 
assuming 80\% polarization for
electrons and 60\% for positrons, an accuracy of
$\Delta \sin^2\theta_{\rm eff} = 1.3 \times 10^{-5}$ can be
achieved in the leptonic final state~\cite{TDR}.

\vspace*{.1cm}
\begin{figure}[htb!]
\setlength{\unitlength}{1cm}
\begin{minipage}{10cm}
\begin{picture}(9,4.5)
\put(-1.,0)
{\includegraphics[width=5cm,height=4.6cm]{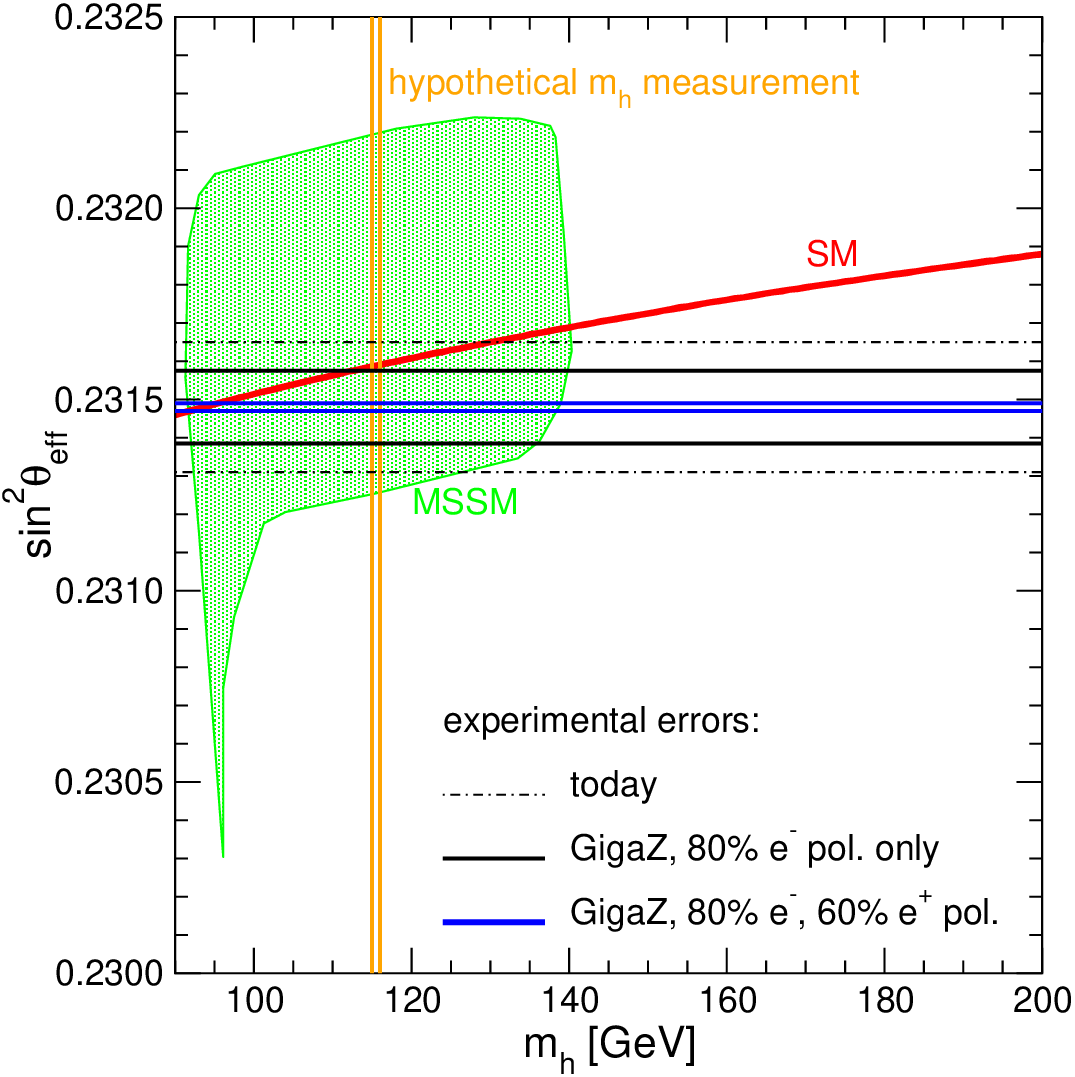}}
\put(4.2,0.)
{\includegraphics[width=5cm,height=4.6cm]{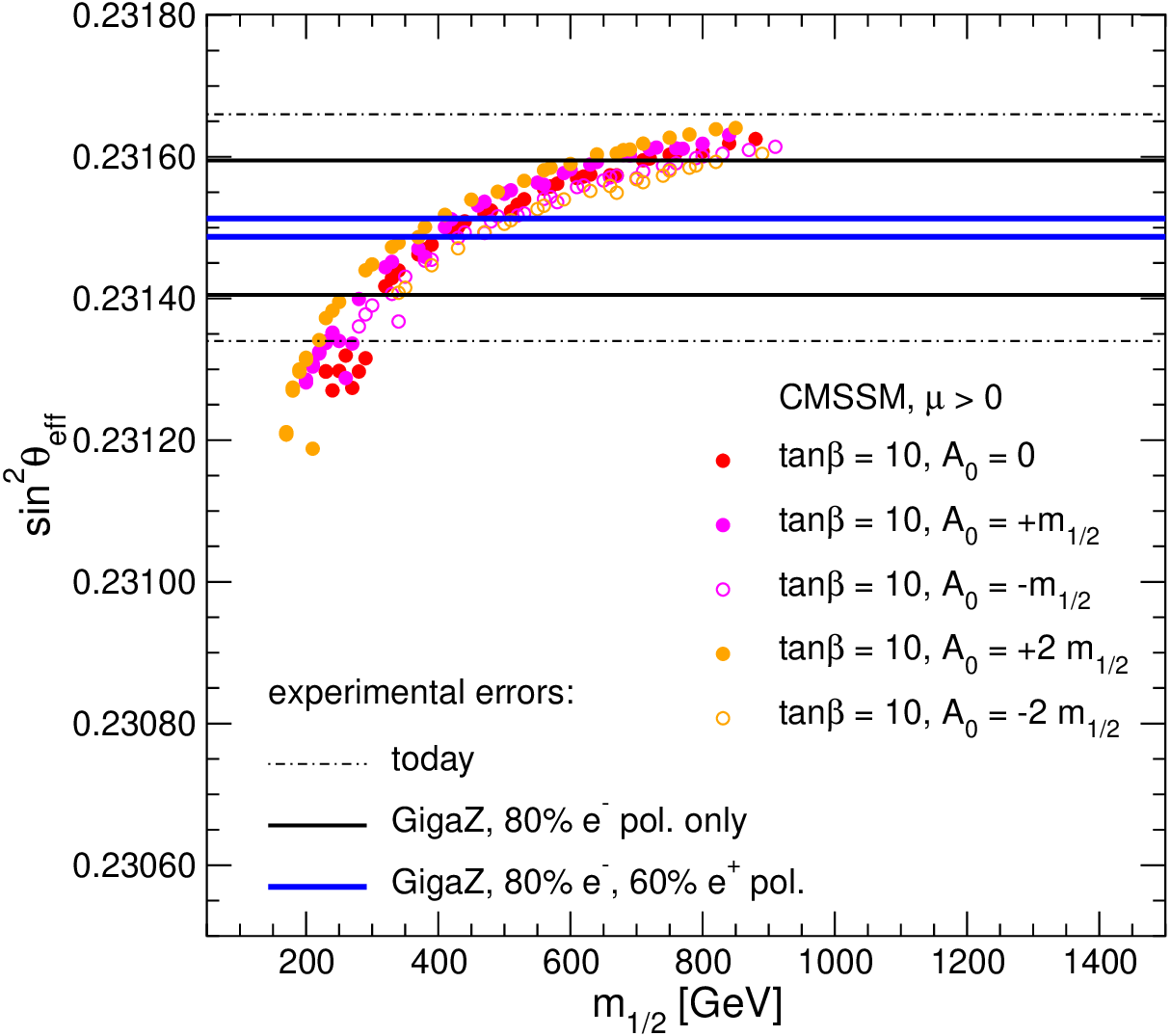}}
\end{picture}
\end{minipage}
\begin{minipage}{7.5cm}
\vspace{-.5cm}
\begin{tabular}{l}
Fig.~5: 
{\it The theoretical predictions for $\sin^2 \theta_{\rm eff}$ in terms}\\ 
{\it of $m_h$, the mass of the Higgs boson in the SM or}\\ 
{\it the mass of the lightest Higgs boson in the MSSM,}\\ 
{\it respectively, are compared with the experimental}\\ 
{\it accuracies obtainable at GigaZ (left panel).}\\
{\it Right panel: The precision measurement of $\sin^2 \theta_{\rm eff}$}\\
{\it yields constraints on
the allowed range for the SUSY}\\ 
{\it mass parameter $m_{1/2}$ in a specific model, the}\\
{\it CMSSM. Experimental constraints from LEP}\\ 
{\it searches and cold-dark-matter searches have been}\\ 
{\it taken into account.}
\end{tabular}
\end{minipage}
\end{figure}

\vspace*{-.3cm}
Because of the gain of about 1 order of magnitude in the accuracy of $\sin^2 \theta_{\rm eff}$,
the bounds on $m_h$ in the SM improve by about 1 order of magnitude, and
the allowed range of $m_{1/2}$ is reduced by a factor of about 5 when using
$(|P_{e^-}|,|P_{e^+}|)=(80\%,60\%)$ instead of $(|P_{e^-}|,|P_{e^+}|)=(80\%,0\%)$.
\subsection{Transversely-polarized beams\\[-.5em]}
With the polarization of both beams, another powerful tool will be
available at the ILC, namely the use of transversely-polarized beams.
These significantly enhance the physics potential
for SM physics as well as for different new-physics
models:  new CP-sensitive observables can be probed
and azimuthal asymmetries can be exploited, which is particularly 
important in SUSY searches for new CP-violating
sources; for further detailed examples, see~\cite{Moortgat-Pick:2005cw}.
These asymmetries
are also sensitive to
new kinds of interactions, e.g.\ spin-2 graviton exchanges in certain
extra-dimension models.  However, both beams have to be polarized,
otherwise all effects from transverse polarization vanish in the limes $m_e\to 0$ 
(suppression $\sim m_e/\sqrt{s}$).

Furthermore, an example from SM physics of an additional benefit 
of having transversely-polarized beams in
testing the electroweak gauge group is the unique access to one specific
triple gauge coupling (TGC).
In the SM the most general parametrization of the gauge-boson
self-interactions leads to 14 complex parameters.
It turns out that for most couplings
longitu\-di\-nally-polarized $e^-$ and $e^+$ beams are sufficient with the exception of
the coupling, $\tilde{h}_{+}={\rm Im}(g^{\rm R}_1+\kappa^{\rm L})/\sqrt{2}$,
which is only accessible with transversely-polarized beams.
Concerning the determination of the other TGC
the gain is of a factor of about 1.8,
when applying both beams longitudinally polarized instead of only polarized electrons.

Transversely polarized beams provide sensitivity to non-standard interactions,
which are not of the current--current type,
such as those mediated by tensor or (pseudo)scalar exchanges. This is the case
even in indirect searches, see Fig.~6. 

\begin{figure}
\begin{minipage}{4.5cm}
\setlength{\unitlength}{1cm}
\begin{picture}(8,3.)
\put(-.5,.0){\includegraphics[width=3.5cm,height=5cm,angle=90]{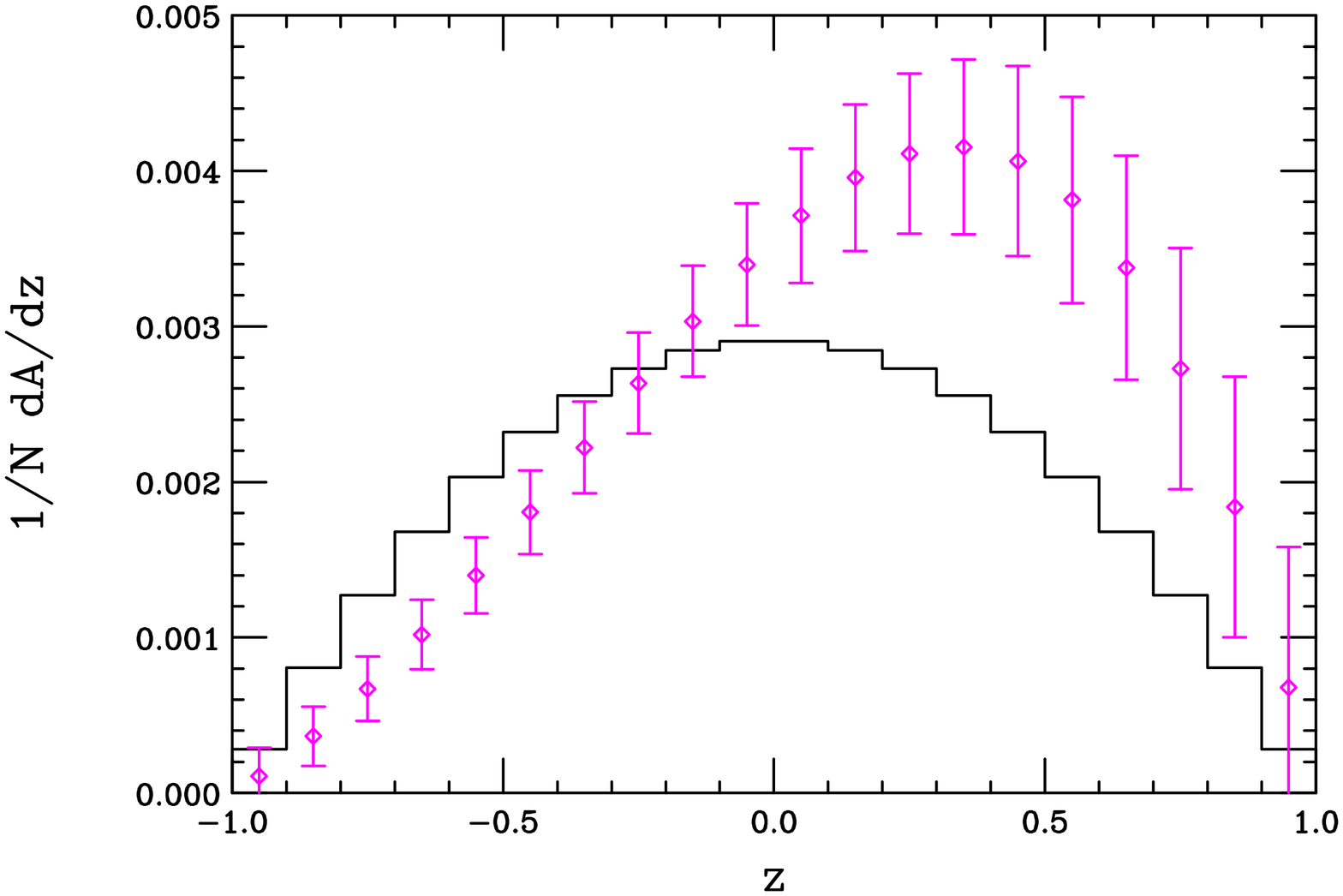}}
\put(1.5,3.){\small ADD}
\put(2.3,1.5){\small SM}
\end{picture}
\end{minipage}\hspace*{.1cm}\vspace*{-.5cm}
\begin{minipage}{11.5cm}
\vspace*{-1cm}
\begin{tabular}{l}
Fig.~6: {\it One representative example is the distinction between
extra dimensions in}\\ {\it the models of Randall--Sundrum (RS) and
Arkani-Hamed--Dimopoulos--Dvali (ADD).}\\ {\it The azimuthal
distributions in the SM as well as the RS--model have to be
absolutely}\\ {\it symmetric, whereas the asymmetric behaviour of the
azimuthal asymmetry in the}\\ {\it ADD-model clearly shows the effects
of the spin-2 graviton. For such a model test}\\ {\it with
transversely-polarized beams, the polarization of both beams is
required.}
\end{tabular}
\end{minipage}
\end{figure}

The possibility to identify new physics even in indirect searches when
applying polarized $e^-$ and $e^+$ beams represents a further step forward in
our understanding of fundamental interactions.

\section{NEWS FROM TECHNICAL ASPECTS OF POLARIZED $\boldmath{e^-}$ AND $\boldmath{e^+}$ 
BEAMS AT THE ILC \\[-.5em]}
\noindent {\bf Positron sources at the ILC}\\
Several possibilities for a positron source for the ILC are under discussion:
{\bf a)} a conventional, non-polarized source, {\bf b)} a helical undulator-based polarized source,
{\bf c)} a laser-based polarized source.

The conventional positron source assumes a primary electron of 6.2 GeV, the nominal 
ILC beam current and bunch structure for the $e^+$ beam generation, and a thick target of about $\ge 4$ 
radiation lengths (r.l.) of tungsten~\cite{floettmann}.
The polarized sources $b)$ and $c)$ use circularly-polarized photons
(generated via undulator radiation or a Compton-backscattering of laser
light) and need only a thin target of about $0.5$ r.l. 

Solution b) needs a high-energy electron beam ($\ge 150$~GeV),
whereas in c) only a few-GeV $e^-$ beam is needed, which can be
generated in a stand-alone linac.  The impact of the linked operation
of an undulator-based source on the overall machine performance is
still under discussion, but recent results show~\cite{daresbury} that
it can be greatly reduced by using an additional low-intensity
electron keep-alive beam.  On the contrary, such polarized sources provide a
much smaller $e^+$ beam divergence, resulting in a large safety margin,
less heat load of the target, a higher capture efficiency and allow an
$e^+$ source for a damping-ring acceptance smaller than for a
conventional $e^+$ source~\cite{floettmann}.

$\bullet$ The proof-of-principle experiment for the undulator-based polarized-positron source
is the currently running project E-166 at SLAC.
It uses the 50 GeV FFTB to generate, via a 1-m long  helical undulator, polarized photons that are
then converted at a thin target into polarized positrons. The polarization of the photons as well
as the positrons will  then be analysed and compared with the theoretical simulations.
Since the photon spectrum, the chosen target material and the thickness are similar to those foreseen for
the possible ILC design, polarized positrons will be produced with the same polarization 
characteristics as expected at the ILC. Already the first run led to a large amount of excellent data
and has shown that undulator radiation provides a feasible, stable $e^+$ source. 
The second run is scheduled for September 2005, and final results are expected
at the end of the year. 

$\bullet$ Helical-undulator prototypes for a specific ILC design
are currently developed under the guidance of the Daresbury and Rutherford 
Laboratories, U.K.
Two designs are discussed: the first device uses
superconducting magnets and the second one a Halbach undulator with permanent magnets.
The choice between the two technologies is foreseen for this year.

$\bullet$ Concerning the laser-based source, a prototype experiment was
ongoing at KEK, which demonstrated that the polarization transfer from the laser to the
circularly polarized photons, as well as from the polarized photons to the positrons, is as expected.\\

\noindent {\bf Polarization measurement and spin manipulations}\\
For both polarized electrons and polarized positrons,
the polarization measurement will be done with Compton
polarimetry. Depending on the final choice of a beam head-on design or a crossing angle design,
the polarimeter could be installed upstream or/and downstream. 
For both designs, a magnet chicane system seems to be very useful:
for an upstream polarimeter, a chicane enables us to retain maximum coverage of the electron detector;
for a downstream polarimeter a chicane is needed to discern between Compton-edge electrons and 
the low-energy disrupted primary electrons. 

The expected polarimeter precision at the ILC is expected to be
$\Delta P_{e^-}/P_{e^-}\sim \Delta P_{e^+}/P_{e^+} \le 0.5\%$, up to
$0.1\%$--$0.2\%$.  For such a polarimeter either a specialized
high-power laser or a conventional laser amplified by a Fabry--Perot
cavity is needed~\cite{lcws05}.  To get even higher precision, $\Delta
P/P < 0.1\%$, the Blondel scheme for polarization
measurement must be used. This scheme requires polarized positrons and, in
particular, that the polarization of the two beams be switched independently.
To keep systematics under control, a fast switching is desired.

$\bullet$ Pulse-to-pulse switching of the positron polarization can be
accomplished by utilizing slow kicker magnets. A pair of dipoles is
turned on between pulse-trains so as to deflect the beam through solenoids
to rotate the spin to the opposite helicity. With such a system, the
change of positron polarization can be made between pulse-trains, which
is fast enough to keep any systematics well under control.

$\bullet$ To provide transversely-polarized beams, we just have to change
the spin rotator settings  ---consisting of two solenoids
and a bend-rotation system, while minimizing the emittance dilution---
just after the damping system.
Such a device will allow us to set the spins at any arbitrary
orientation by the
time they reach the interaction region.
Longitudinal Compton polarimeters  can monitor that the longitudinal polarization stays close to zero.
Applying periodically ($\sim$~every 1--2 days), a precision
of $\Delta P^{\rm T}/P^{\rm T}\sim 1\%$ should be achieved.

\section{CONCLUSIONS AND OUTLOOK \\[-.5em]}

In Ref.~\cite{Moortgat-Pick:2005cw} many examples within the Standard Model, as well as
from numerous models beyond it have demonstrated in detail that
simultaneous polarization of the $e^-$ and $e^+$ beams will provide a very
efficient tool for direct as well as indirect searches for new
physics. The option of polarizing both beams provides a powerful tool for
studying new physics  at the ILC, such as discovering new particles, 
analysing signals model-independently and precisely resolving the underlying model.
The polarization of both beams therefore serves as a superior experimental
tool to face the (expected and unforeseen) challenges of possible new
physics, as well as to make the
high-precision studies of SM processes possible.

One should keep in mind that the given
examples, however, by no means exhaust the whole phenomenology of
simultaneously polarized electron and positron beams. Many studies
are ongoing, and further ideas for the exploitation of both beams
polarized are coming up.

Techniques and engineering designs for a polarized-positron source
are well advanced. Potential challenges concerning
luminosity, commissioning and operating issues appear to be under control, e.g.\ with an 
additional low-energy keep-alive beam. Therefore,
consideration can now be given to including a polarized-positron
source already in the baseline design.

\begin{acknowledgements}
\vspace*{-.2cm}
I would like to thank all authors of \cite{Moortgat-Pick:2005cw} for their active and
lively collaboration. I am particularly grateful to Sabine Riemann and 
Desmond Barber, Vinod
Bharadwaj, Jym Clendenin, Eckhard Elsen, Klaus Fl\"ottmann,
Hans Fraas, Stefan Hesselbach, Jan Kalinowski,
Uli Martyn, Klaus Moenig, Uriel Nauenberg, Tsunehiko Omori, Rainer
Pitthan, Peter Schueler, John Sheppard, Alessandro Variola, 
Nick Walker, Georg Weiglein, Mike Woods and 
Peter Zerwas, for
many helpful and encouraging discussions. 
I am strongly indebted to Per Osland and Nello Paver  
for their great support in editing the physics part of the polarization 
report.  
\end{acknowledgements}

\end{document}